# Stochastic Power Grid Analysis Considering Process Variations

Praveen Ghanta*    Sarma Vrudhula*    Rajendran Panda**    Janet Wang*
*University of Arizona, Tucson, USA, {ghanta,sarma,wml}@ece.arizona.edu
**FreeScale Semiconductor Inc., Austin, USA, Rajendran.Panda@freescale.com

## Abstract

*In this paper, we investigate the impact of interconnect and device process variations on voltage fluctuations in power grids. We consider random variations in the power grid's electrical parameters as spatial stochastic processes and propose a new and efficient method to compute the stochastic voltage response of the power grid. Our approach provides an explicit analytical representation of the stochastic voltage response using orthogonal polynomials in a Hilbert space. The approach has been implemented in a prototype software called OPERA (Orthogonal Polynomial Expansions for Response Analysis). Use of OPERA on industrial power grids demonstrated speed-ups of up to two orders of magnitude. The results also show a significant variation of about ± 35% in the nominal voltage drops at various nodes of the power grids and demonstrate the need for variation-aware power grid analysis.*

## 1. Introduction

One of the most difficult and important challenges posed by deep sub 100 nm IC technology is the increasing uncertainty in the performance of CMOS circuits due to variations in the fabrication process [20, 21]. Examples of the parametric variations include variations in doping profiles, materials, interconnect width ($W$) and thickness ($T$), device characteristics like the effective channel length ($L_{eff}$) etc. Physical parameters like $W$, $T$ of the interconnects and $L_{eff}$ of the MOS devices vary significantly [21] with strong intra-die (across die) and inter-die (across wafer) components. These physical variations lead to substantial variations in the electrical parameters viz., conductance, capacitance, inductance, threshold voltages, leakage currents *etc.*, of the CMOS devices and interconnects.

An important aspect of the IC design process is the integrity of the power grid. The exponential increases in transistor density has resulted in huge power distribution networks carrying large transient currents, that result in significant voltage drops (IR and $L\frac{di}{dt}$) in the voltage levels at the power grid nodes. This combined with continuous reduction in supply voltages ($\leq 1.2V$ for the 90 nm process) make the voltage drops critical as they can severely impact the functionality and the performance of the ICs. Process variations in the power grid conductors and onchip CMOS devices of the functional blocks can have a significant detrimental impact on the voltage levels at the power grid nodes. Given the critical dependence of the sub 100nm ICs on the $V_{DD}$ voltage variations, its important to characterize the effects of process variations on the voltage levels at the power grid nodes.

An established body of literature is available on the analysis and optimization aspects of power grids [1–5, 7, 9–13]. Some of the early work on power grid analysis used DC analysis to obtain the IR drops [1–3]. The variations in current profiles of the functional blocks are often obtained by simulation [1, 7, 9, 10]. The authors of [4] describe a multi-grid approach where coarser meshes are first solved using fast PDE solvers and the results are extrapolated to solve the original mesh. An efficient hierarchical power grid analysis technique employing sparsification based on integer linear programming is described in [5] and [6] presents a linear time algorithm based on random walks that is efficient for incremental localized analysis. The analysis of power grids has only been recently expanded to account for process variations. Assuming normal and lognormal distributions for the device threshold voltages and leakage currents, the authors of [12, 13], present a procedure to compute the mean and bounds on the variance of the voltage drops on a power grid. In [11], the voltage response of a power grid is expressed as a convolution of the grid impulse response and load currents. Their approach to account for variability is to view the load currents of the functional blocks (inputs to the power grid system) as random variables due to the large space of input patterns. Based on this, they compute the mean and variance of the voltage response of the grid.

## 2. Our Contributions

In this paper, we propose a new approach [18] to account for the impact of process variations in the analysis of power grids. Until recently, stochastic analysis of systems generally meant that the system inputs were stochastic but the system itself was deterministic with deterministic parameters. The approach to be described here addresses the important case when the system parameters are also stochastic quantities. For such a case, the physical system itself is an outcome of a stochastic process. Our analysis, thus is completely different from all the previous power grid analyses. In this work, due to the manufacturing variations in the interconnect width $W$, thickness $T$ and the device channel length $L_{eff}$, the electrical parameters in the power grid (R,L,C) are modeled as continuous parameter (spatial) stochastic processes. The key development here is an expansion for the stochastic voltage response as an infinite series of orthogonal polynomials of random variables in an infinite dimensional Hilbert space. The expansion can be *optimally* truncated to any order depending on the available computational resources and accuracy requirements. With an explicit analytical representation of the voltage response available in terms of the random variables, moments and probability density functions of voltage can be directly computed. This provides an attractive alternative to the computationally expensive Monte Carlo simulations. Our approach has been implemented in a prototype software called OPERA.

In Section 2, we discuss the core contributions of our work. Section 3 contains the problem definition. In Section 4, we discuss the theoretical foundations of our approach. In Section 5, we demonstrate our method through an example followed by the discussion of a special case. We also provide a brief discussion on the implementation issues. The experimental results are shown in Section 6 and conclusions are presented in Section 7.

## 3. Problem Statement

We consider an RC model of the power grid with a mesh structure in which the metal interconnects and the vias are modeled as passive RC networks. Power sources provide connections from the external supply to the grid and are modeled as ideal voltage sources between the grid nodes and the ground. The functional blocks (logic gates/latches) that are distributed across the chip act as power drains. They are modeled as known transient current sources between the grid nodes and the ground, in parallel with the non-switching capacitances of the functional blocks. The transient current source profiles are obtained from simulation of the functional blocks at a full supply voltage for a large sequence of random input vectors. The package pin contacts that provide the power supply connection are modeled as resistances in series with the supply sources.

The power grid model can be described in the Laplace domain by the MNA equation

$$(G + sC) x(s) = U(s) \quad (1)$$

where $(G+sC)$ is the coefficient matrix and $x(s)$ is the response to be determined. $U(s) = (i(s), G_1 V_{DD})^T$ is the known excitation; vector $i(s)$ represents the functional block drain currents; $G_1$ is a diagonal matrix that consists of non-zero elements at those nodes where $V_{DD}$ sources are connected. Almost all the conductance of the power grid comes from the metal interconnects while they contribute only about 5% of the grid capacitance [8]. The majority of the capacitance contribution of the power grid comes from the non-switching load capacitances of the gates in the functional blocks. The load capacitance of the gates in turn comes from two major sources - the gate ($C_{GS}$ and $C_{GD}$) capacitance of the gates they drive and the source/drain diffusion capacitances ($C_{DB}$ and $C_{SB}$); the two sources contribute almost equally to the grid capacitance in the present technologies [8, 15]. The drain currents $i(s)$ consist of MOS drain currents and the gate oxide and subthreshold leakage currents.

The circuit parameters $G$ and $C$ depend on the grid interconnect and device parameters such as the metal thickness ($T$), metal width ($W$), channel length ($L_{eff}$) etc.. In addition, the MOS drain currents $i(s)$ in the excitation $U(s)$ vary significantly with changes in $L_{eff}$. In the presence of random process variations, these physical characteristics of the interconnects and devices, and hence their electrical characteristics are modeled as spatial stochastic processes (they vary randomly and spatially across a die for intra-die variations and across a wafer for inter-die variations).

In the present work, we assume that the conductance $G$ of the grid varies with parameters $W$ and $T$. We ignore variations in $C$ due to $W$ and $T$. We assume that the gate capacitance contributes about 40% of the total grid capacitance and that this gate capacitance varies with $L_{eff}$. We have from SPICE models $C_{gate} \propto W_{eff} L_{eff} C_{ox}$ where $W_{eff}, C_{ox}$ represent the effective width and gate oxide capacitance of a MOS transistor. The drain currents $i(s)$ that consist of MOS drain currents and leakage currents are known to vary significantly with $L_{eff}$. We consider only the inter-die variations in this work and hence the variations in $G$ and $C$ are fixed for any one single die. Thus, they are modeled as random variables and not as spatial random processes.

Consider the manufacturing process. Every trial of the manufacturing process (trial denotes the fabrication of a die or a wafer) results in a different value for $W$, $T$ and $L_{eff}$ of the interconnects and devices. These parameters are thus functions that map one point in the manufacturing sample space to some real value. Thus, $W$, $T$ and $L_{eff}$ are random variables over the manufacturing sample space. In general, let $\Omega$ denote the manufacturing sample space. For $\omega \in \Omega$, $\xi_i : \omega \to R$ denotes a random variable. Let $\boldsymbol{\xi}(\omega) = (\xi_1(\omega), \ldots, \xi_n(\omega))$ denote a vector of $n$ such random variables. Let $\Theta : \Omega \to R$ denote the vector space of all random variables $\xi_i$. In the presence of process variations the MNA equations for the interconnect can be expressed as:

$$( G(\boldsymbol{\xi}(\omega)) + s\, C(\boldsymbol{\xi}(\omega)) ) \, x(s, \boldsymbol{\xi}(\omega)) = U(s, \boldsymbol{\xi}(\omega)) \quad (2)$$

Equation (2) is known as a stochastic differential equation as the operator $(G(\boldsymbol{\xi}(\omega)) + sC(\boldsymbol{\xi}(\omega)))$ is a stochastic process dependent on a random variable $\boldsymbol{\xi}$ and the deterministic Laplace parameter $s$. Further, the excitation vector $U(s, \boldsymbol{\xi}(\omega))$ has both deterministic and random components. For each manufacturing outcome $\omega$ and for each corresponding value of the parameter $\xi(\omega)$, $x(s, \boldsymbol{\xi}(\omega))$ denotes the fixed response of the system for that particular manufacturing outcome. In the next section, we discuss the theoretical foundations of our proposed approach to compute the response $x(s, \boldsymbol{\xi}(\omega))$.

## 4. Proposed Approach

The approach presented here is based on representing the stochastic voltage response $x(s, \boldsymbol{\xi}(\omega))$ of the power grid as an infinite series of orthogonal polynomials in an infinite dimensional Hilbert space of random variables. We assume that $x(s, \boldsymbol{\xi}(\omega))$ is a second order process, i.e. all the random variables have finite variances. *Without loss of generality, we also assume that the variables $\boldsymbol{\xi}(\omega)$ are random variables with zero mean and unit variance. Given a random variable $W$ with mean $W_\mu$ and standard deviation $W_\sigma$, $W$ can always be expressed in a normalized form as $W = W_\mu + W_\sigma \xi_W$, where $\xi_W$ is a random variable with zero mean and unit variance.*

For simplicity, henceforth we will write $\boldsymbol{\xi}(\omega)$ as $\boldsymbol{\xi}$. To explain our approach, we state a few well known facts from the theory of orthonormal expansions [24].

- Let $V$ be an inner product space, with the inner product denoted by $\langle \cdot, \cdot \rangle$. For non-zero $x, y \in V$, $x$ and $y$ are *orthogonal* if $\langle x, y \rangle = 0$. They are *orthonormal* if $\|x\| = \|y\| = 1$.
- An complete inner product space $H$ (i.e. one in which every Cauchy sequence converges) is called a Hilbert space.
- An infinite family of orthonormal vectors, $\{\phi\}_{k=1}^{\infty}$, in a Hilbert space is called an *orthonormal basis* if it is a maximal set of mutually orthonormal vectors.
- If $\{\phi\}_{k=1}^{\infty}$ is a orthonormal basis of a Hilbert space $H$, then the infinite series $\sum_{k=1}^{\infty} \langle x, \phi_k \rangle \phi_k$ converges in norm to $x$.

We now return to the problem of representing the stochastic voltage response $x(s, \boldsymbol{\xi})$. From Section 3, we know that the space $\Theta : \{\Omega \to R\}$ denotes the infinite dimensional vector space of mappings, each mapping representing a random variable $\boldsymbol{\xi}(\omega)$. Let $P$ be the probability measure on the sample space $\Omega$ i.e., the random variables $\boldsymbol{\xi}$ have the probability density function $P$. This space of mappings forms a Hilbert space $\mathcal{H}_\Theta$, where the inner product of any two mappings is the expected value of their product under $P$. That is,

$$< \boldsymbol{\xi}_m, \boldsymbol{\xi}_n > = E(\boldsymbol{\xi}_m, \boldsymbol{\xi}_n) = \int_\Omega \boldsymbol{\xi}_m \, \boldsymbol{\xi}_n \, dP \quad (3)$$

The stochastic voltage response $x(s, \boldsymbol{\xi})$ is an element of $\Theta$. Therefore, if we can find a orthonormal basis $\{\gamma_1(\boldsymbol{\xi}), \gamma_2(\boldsymbol{\xi}), \ldots\}$ for $\mathcal{H}_\Theta$, then $x(s, \boldsymbol{\xi})$ can be represented by

$$x(s, \boldsymbol{\xi}) = \sum_{n=0}^{\infty} a_n(s) \, \gamma_n(\boldsymbol{\xi}) \quad (4)$$

Equation (4) is a general representation of the stochastic voltage response process $x(s,\boldsymbol{\xi})$ in terms of the orthonormal basis functions of random variables $\boldsymbol{\xi}$. The primary task now is to identify the orthonormal basis $\{\gamma_1(\boldsymbol{\xi}), \gamma_2(\boldsymbol{\xi}), \ldots\}$ in $\boldsymbol{\xi}$. One such orthonormal basis is the set of Hermite polynomials of all orders. This basis is valid for any second order process. However, other types of polynomials can also serve as orthonormal bases. The orthogonality is defined w.r.t the norm in Equation (3) which is dependent on the probability density function $P$ of $\boldsymbol{\xi}$. So, for different probability distributions of the random variables, different orthonormal basis sets need to be identified. *The well established Askey scheme of polynomials [19] helps us identify the orthogonal polynomials for different probability density functions. For example, if the underlying random variables are Gaussian or lognormal, then the best choice (in terms of speed of convergence) would be Hermite polynomials. Similarly, for Gamma, Beta and Uniform random variables, the best choices would be Laguerre, Jacobi and Legendre polynomials respectively.*

For the sake of demonstrating our approach, we consider a power grid subjected to inter-die Gaussian variations in interconnect width ($W$), thickness ($T$) and device channel length ($L_{eff}$). The methodology described here can be easily extended to consider other probability distributions and any number of variables.

### 4.1. Gaussian Random Variables

Let $\boldsymbol{\xi} = \{\xi_1, \xi_2, \xi_3, \ldots, \xi_n\} \in \Theta$ be zero mean orthonormal Gaussian random variables. Hermite polynomials of all orders in $\boldsymbol{\xi}$ form an orthonormal basis for $\mathcal{H}_\Theta$. They are defined by

$$H_p(\{\alpha_1, \alpha_2, \cdots, \alpha_p\}) = (-1)^p e^{\frac{1}{2}\boldsymbol{\xi}^t\boldsymbol{\xi}} \frac{\partial^p}{\partial \alpha_1 \partial \alpha_2 \cdots \partial \alpha_p} e^{-\frac{1}{2}\boldsymbol{\xi}^t\boldsymbol{\xi}}, \quad (5)$$

where $\{\alpha_i\}$ are any set of $p$ variables chosen from the set $\{\xi_1, \xi_2, \cdots, \xi_n\}$ with repetitions. Since $p$ variables can be chosen from a set of $n$ variables in $M = (p+n-1)!/p!(n-1)!$ ways, the number of Hermite polynomials of degree $p$ is given by M. As an example, Hermite polynomials in $\{\xi_1, \xi_2\}$ of order 0, 1, 2 can be found as follows.

$$\begin{aligned}
\text{order 0:} \quad & H_0(\{\}) = 1, \\
\text{order 1:} \quad & H_1(\xi_1) = \xi_1,\ H_1(\xi_2) = \xi_2, \\
\text{order 2:} \quad & H_2(\xi_1, \xi_1) = \xi_1^2 - 1,\ H_2(\xi_1, \xi_2) = \xi_1 \xi_2, \\
& H_2(\xi_2, \xi_2) = \xi_2^2 - 1
\end{aligned} \quad (6)$$

Let $\{\xi_1, \xi_2, \xi_3, \ldots\}$ denote an infinite set of zero mean orthonormal Gaussian random variables. Then the stochastic response can be represented as a convergent infinite series expansion of Hermite polynomials as [18]

$$\begin{aligned}
x(s,\boldsymbol{\xi}) &= b_0(s) H_0 + \sum_{i_1=1}^{\infty} b_{i_1}(s) H_1(\xi_{i_1}) \\
&+ \sum_{i_1=1}^{\infty}\sum_{i_2=1}^{i_1} b_{i_1 i_2}(s) H_2(\xi_{i_1}, \xi_{i_2}) \\
&+ \sum_{i_1=1}^{\infty}\sum_{i_2=1}^{i_1}\sum_{i_3=1}^{i_2} b_{i_1 i_2 i_3}(s) H_3(\xi_{i_1}, \xi_{i_2}, \xi_{i_3}) \\
&\vdots
\end{aligned} \quad (7)$$

**Note:** A closer look at Equations (6), (7),(4) shows that Equation (4) is simply a rearrangement of the terms in Equation (7). The remaining task then is to determine the coefficients $\{a_i\}$ of the expansion given in Equation (4).

The expansion shown in Equation (7) called the *Homogeneous Chaos* was developed by Weiner [23]. The celebrated result of Cameron and Martin [17] extended the expansions to general function spaces, and is now referred to as *Polynomial Chaos*. Ghanem and Spanos [18] developed applications of these results to the study of systems with stochastic parameters.

### 4.2. Evaluation of the Coefficients

Though the number of random variables $\boldsymbol{\xi}$ are finite, the order $p$ and hence the number of terms $N+1$ in the stochastic response expansion are infinite. For practical computation purposes, the infinite series needs to be projected onto a finite space, i.e. truncated. The general approach is to limit the response expansion to a finite order $p$, which in turn determines the accuracy. Often, a second order ($p=2$) or third order ($p=3$) expansion is sufficient. If there are $n$ random variables, the response expansion obtained by limiting the order to $p$ would be

$$x(s,\boldsymbol{\xi}) = \sum_{i=0}^{N} a_i(s)\, \gamma_i(\boldsymbol{\xi}) \quad (8)$$

where $N = \sum_{k=0}^{p} \binom{n-1+k}{k}$.

The error due to truncation is given by

$$\Delta_p(s,\boldsymbol{\xi}) = (G(\boldsymbol{\xi}) + s\, C(\boldsymbol{\xi}))\, x(s,\boldsymbol{\xi}) - U(s,\boldsymbol{\xi}) \quad (9)$$

Once we have the truncated expansion from Equation (8), we need to evaluate the *best* deterministic coefficients $\{a_i\}$ that result in the best minimization of the truncation error. We follow the principle of orthogonality which states that the best approximation of the response $x(s,\boldsymbol{\xi})$ is one in which the truncation error $\Delta_p(s,\boldsymbol{\xi})$ is orthogonal to the approximation. The application of the principle of orthogonality to obtain a finite projection is known as the Galerkin method. The orthogonality is defined w.r.t to some norm, which in this case is given by $<\Delta_p(s,\boldsymbol{\xi}), \gamma_i(\boldsymbol{\xi})>$. Thus, to obtain the deterministic coefficients we need to solve the system of equations given by

$$<\Delta_p(s,\boldsymbol{\xi}), \gamma_i(\boldsymbol{\xi})> = 0,\ i = 0,1,2,\cdots N \quad (10)$$

## 5. Illustration

As an example to illustrate the methods, we consider a power grid subjected to random process variations where $\xi_W$, $\xi_T$, $\xi_L$ denote the **normalized** variations in width, thickness and the effective device channel length respectively. Then

$$\xi_W = \frac{\Delta W}{W},\ \xi_T = \frac{\Delta T}{T},\ \xi_L = \frac{\Delta L_{eff}}{L_{eff}} \quad (11)$$

Without loss of generality, we assume that they are uncorrelated Gaussian random variables. Certainly if they were not, given their covariance matrix, they can always be transformed into a set of uncorrelated random variables by an orthogonal transformation technique like principal component analysis [16]. As discussed earlier, the $G$ matrix depends on $\xi_W$ and $\xi_T$ and the $C$ matrix depends on $\xi_L$. We use a linear model to capture the dependence of $G$ and $C$ on these random variables in accordance with the models in contemporary literature [20]. But there are no limitations on the specific model to be chosen. The MNA equation for the grid system is

given by:
$$(G(\xi)+sC(\xi))\,x(s,\xi) = U(s,\xi) \tag{12}$$

where $\xi = (\xi_W, \xi_T, \xi_L)$ is the random variable vector and the perturbed matrices $G$ and $C$ and $U(s)$ are given by

$$\begin{aligned}
G(\xi) &= G_a(s)+G_b(s)\xi_W+G_c(s)\xi_T \\
C(\xi) &= C_a(s)+C_c(s)\xi_L \\
U(s,\xi) &= U_a(s)+U_b(s)\xi_W+U_d(s)\xi_T+U_c(s)\xi_L
\end{aligned} \tag{13}$$

$G_a, C_a, U_a$ are the mean matrices and $G_b, G_c, C_c, U_b, U_c, U_d$ represent the perturbation matrices of $G, C, U(s)$ w.r.t $W, T, L_{eff}$.

Consider Equation (13) and the physical definition for conductance per unit length $G = \frac{WT}{\rho}$. We can observe that for the linear model in Equation (13), $G_b$ and $G_c$ are same as $G_a$ *scaled by some constants*. So, we have $G_b = dG_a$ and $G_c = eG_a$ where $e$ and $d$ are some constants. Since the scaled sum of two independent Gaussian variables $\xi_W$ and $\xi_T$ ($d\xi_W + e\xi_T$) is a Gaussian variable with calcuable mean and variance, matrix $G$ and hence $U(s,\xi)$ can be re-represented using a single normalized variable $\xi_G$ as

$$\begin{aligned}
G(\xi) &= G_a(s)+G_g(s)\xi_G, \\
U(\xi) &= U_a(s)+U_g(s)\xi_G+U_c(s)\xi_L \\
C(\xi) &= C_a(s)+C_c(s)\xi_L
\end{aligned} \tag{14}$$

$\xi$ is now given by $\xi = (\xi_G, \xi_L)^T$. Using Hermite polynomials as the basis, we can expand the response $x(s,\xi)$ using a second order expansion ($p=2$) from Equation (7) as

$$\begin{aligned}
x(s,\xi) &= a_0(s)+a_1(s)\xi_G+a_2(s)\xi_L+a_3(s)(\xi_G^2-1) \\
&+ a_4(s)(\xi_G\xi_L)+a_5(s)(\xi_L^2-1).
\end{aligned} \tag{15}$$

where $a_i(s)$ is a vector. The coefficient vectors $\{a_i\}$ need to be determined.

Following the orthogonal truncation method from Section 4.2, we can define the error as in (9).

$$\Delta_p(s,\xi) = (G(\xi)+sC(\xi))\,x(s,\xi) - U(s,\xi) \tag{16}$$

Coefficient vectors $\{a_i(s)\}$ are obtained by solving (see Equation (10))

$$\langle \Delta_p(s,\xi), \gamma_j(\xi) \rangle = 0 \quad \text{for } j=0,1,2,\ldots,N \tag{17}$$

The inner product $\langle \Sigma_p(s,\xi), \gamma_j \rangle$ is defined as

$$\begin{aligned}
&\langle \Delta_p(s,\xi), \gamma_j(\xi) \rangle \\
&= \int_{-\infty}^{+\infty}\int_{-\infty}^{+\infty} \Delta_p(s,\xi)\gamma_j(\xi)W(\xi)\,d\xi_G\,d\xi_L = 0,
\end{aligned} \tag{18}$$

where $W(\xi)$ is the bivariate Gaussian probability density function.

Thus for each $j = 0,1,2,\cdots,N$, Equation (17) results in a linear system of $(N+1)$ equations to solve for the deterministic coefficients vectors $\{a_i\}$ represented by the vector $a(s)$.

$$(\tilde{G}+s\tilde{C})\,a(s) = \tilde{U}(s), \tag{19}$$

where

$$\tilde{G} = \begin{bmatrix} G_a & G_g & 0 & 0 & 0 & 0 \\ G_g & G_a & 0 & 2G_g & 0 & 0 \\ 0 & 0 & G_a & 0 & G_g & 0 \\ 0 & 2G_g & 0 & 2G_a & 0 & 0 \\ 0 & 0 & G_g & 0 & G_a & 0 \\ 0 & 0 & 0 & 0 & 0 & 2G_a \end{bmatrix} \tag{20}$$

$$\tilde{C} = \begin{bmatrix} C_a & 0 & C_c & 0 & 0 & 0 \\ 0 & C_a & 0 & 2C_b & C_c & 0 \\ C_c & 0 & C_a & 0 & 0 & 2C_c \\ 0 & 0 & 0 & 2C_a & 0 & 0 \\ 0 & C_c & 0 & 0 & C_a & 0 \\ 0 & 0 & 2C_c & 0 & 0 & 2C_a \end{bmatrix} \tag{21}$$

Matrix $\tilde{U}(s)$ is given by

$$\tilde{U}(s) = (U_a(s),\ U_g(s),\ U_c(s),\ 0,\ 0,\ 0)^T \tag{22}$$

Now, we can solve Equation (19) numerically to obtain the coefficient vector $a(s)$. Once the vector $a(s)$ is obtained, we have an explicit expression for the circuit response $x(s,\xi)$ in terms of $\xi$ given by Equation (15). With this explicit expression, we can obtain the moments of the stochastic voltage response at any node of the power grid as follows:

$$\begin{aligned}
\text{Mean}(x(t,\xi)) &= a_0(t) \\
\text{Var}(x(t,\xi)) &= (a_1(t))^2\,\text{Var}(\xi_G)+(a_2(t))^2\,\text{Var}(\xi_L) \\
&+ (a_3(t))^2\,\text{Var}(\xi_G^2-1)+(a_4(t))^2\,\text{Var}(\xi_G)\,\text{Var}(\xi_L) \\
&+ (a_5(t))^2\,\text{Var}(\xi_L^2-1) \\
\text{Var}(x(t,\xi)) &= (a_1(t))^2+(a_2(t))^2+(a_3(t))^2\,2+(a_4(t))^2 \\
&+ (a_5(t))^2\,2
\end{aligned} \tag{23}$$

To obtain higher order moments for $x(t,\xi)$, we can use the equality $\mathbf{E}(x^n(t,\xi)) = <x^{n-1}(t,\xi), x(t,\xi)>$, provided that $x(t,\xi)$ has an accurate representation using an expansion of sufficiently large order $p$. Once the higher order moments are obtained, expansions like Gram-Charlier series or Edgeworth series could be used to obtain the probability density function of $x(t,\xi)$ directly.

### 5.1. Special Case

As a special case, let's consider only the variations in the drain currents $U(s)$, *i.e.* the R.H.S of the MNA analysis Equation (1) say due to threshold variations. $U(s)$ consists of MOS drain currents and leakage currents; the latter are known to vary exponentially with the threshold voltages. Our problem then becomes similar to the one addressed by Ferzli and Najm [12, 13]. We then have,

$$(G+sC)x(s,\xi) = U(s,\xi) \tag{24}$$

where $\xi$ represents the threshold variations. If a normal distribution for $V_{th}$ is assumed, then the distribution for the leakage currents becomes lognormal. To consider intra-die variations in $V_{th}$, let's divide the chip in to a finite number of regions say 2 for this example. Let's assume that $\xi = (\xi_1,\xi_2)$ are the normalized uncorrelated Gaussian random variables that represent $V_{th}$ variations for the two chip regions. *Then, the currents $U(s,\xi)$ can always be expressed using an orthogonal Hermite polynomial basis [25] in $\xi_1,\xi_2$ to any required order ($p$) of accuracy, say $p=2$ for this case*. Since, the R.H.S of Equation (24) is stochastic, the response of the grid becomes stochastic as well and can be represented using the Hermite basis. We thus have,

$$\begin{aligned}
x(s,\xi) &= x_0(s)+x_1(s)\xi_1+x_2(s)\xi_2+x_3(s)(\xi_1^2-1) \\
&+ x_4(s)(\xi_1\xi_2)+x_5(s)(\xi_2^2-1) \\
U(s,\xi) &= U_0(s)+U_1(s)\xi_1+U_2(s)\xi_2+U_3(s)(\xi_1^2-1) \\
&+ U_4(s)(\xi_1\xi_2)+U_5(s)(\xi_2^2-1)
\end{aligned} \tag{25, 26}$$

Following the error minimization procedure illustrated in Section 5, our analysis simply translates to solving *independent equations* of the form for $n = 0, 1, \ldots, 5$

$$(G + sC) x_n(s) = U_n(s) \quad (27)$$

All we need is a single $LU$ factorization of the original matrix $(G + sC)$ and then repeated solves for different values of the R.H.S. Using the formulae from Equation (23), we can directly compute the mean, variance and other higher order moments unlike [12, 13] which can calculate only some bounds for the variance.

### 5.2. Implementation Issues

The procedure for obtaining the voltage response has been implemented in a prototype software called OPERA. One of the primary issues involved is the computational complexity of our approach. Solving the system of Equations (19) is computationally the most intensive step in our approach. The complexity of that step is dependent on the length of the vector $a(s)$ which varies as $O(r^p)$, where $p$ is the order of the expansion and $r$ is the number of random variables. We found an order 2/order 3 expansion to be sufficiently accurate for variational power grid analysis considering realistic bounds for maximum variability in the grid interconnect and device parameters. Also, the matrices $\tilde{G}$ and $\tilde{C}$ are very sparse and they have been observed to become increasingly sparser with an increase in the order of the expansion ($p$) or the number of random variables ($r$).

Further, computational complexity of OPERA can be significantly reduced by efficient techniques like model order reduction (MOR) [14], multi grid analysis [4] and iterative block solvers with appropriate pre-conditioners [18]. MOR techniques can be used as the power grid node voltages in the top layers and their moments w.r.t $\xi$ are typically of no interest to the designer. Model order related stability issues have been addressed in a number of literatures and any existing stability technique can be incorporated in our method.

### 6. Experimental Results

OPERA has been verified for many industrial power grids considering order 2/order 3 expansions for representing the stochastic voltage response. Results for a few grids are presented here. As discussed earlier, we assume a Gaussian distribution for the variables $W$, $T$ and $L_{eff}$ for all the grids. We consider a linear or a first order model for the variations in $G$, $C$ and the drain currents $i(s)$. Leakage currents are known to vary exponentially with $L_{eff}$ which may suggest a higher order expansion for $U(s)$ in $L_{eff}$. But for the purposes of this paper, we consider a linear expansion itself as they constitute only 5% of the total currents in the current CMOS technologies [22].

*Table 1 shows the results for the transient analysis of 7 industrial grids for maximum $3\sigma$ variations of* 20% *in* $\xi_W$, 15% *in* $\xi_T$ *(hence* 25% *in* $\xi_G$*) and* 20% *in* $\xi_L$. *Note that we are considering the inter-die variations and that only* 40% *of the capacitance varies with* $L_{eff}$ as discussed earlier. A fixed time step was used in carrying out the transient analysis and an order 2 expansion was used for representing the stochastic response. Comparison of results from OPERA with Monte Carlo simulations (1000 samples for each case) has been done for all the grids and the results are reported in Table 1. The average and the maximum errors in obtaining the mean ($\mu$) and variance ($\sigma$) of the voltage response from OPERA compared to Monte Carlo simulations are shown in the table. The errors reported in Table 1 are for data obtained from simulation across all nodes and all time points of the transient simulation of the grid. We can see from the table that OPERA demonstrates good accuracy in determining the mean and variance of the voltage response. The accuracy in obtaining the variance by OPERA can be increased further by increasing the order of the expansion.

The drain current profiles used for the transient analysis of the power grid were such that the peak drop in the voltage at any grid node was less than 10 % of the $V_{DD}$. Under such conditions, it was observed that for all grids the mean voltage drops at the grid nodes ($\mu$) with variations was more or less the same as the nominal voltage drops ($\mu_0$) with out variations. And the difference between the two drops ($\mu - \mu_0$) was negligible when expressed as a fraction of % of $V_{DD}$. *However on average for each grid, the $\pm 3\sigma$ variation in the voltage drops at the grid nodes, was about $\pm 35$ % of $\mu_0$, where $\mu_0$ are the nominal voltage drops with out variations*. This strongly supports the necessity of considering the effects of process variations on power grids.

For the power grid with 19,181 nodes, we plotted the distribution of the voltage response from OPERA and Monte Carlo simulations w.r.t to the variations in $\xi_W$, $\xi_T$ and $\xi_L$ at arbitrarily selected nodes in the power grid. Figures 1, 2 show the voltage distribution at a select node from Monte Carlo simulations and OPERA.

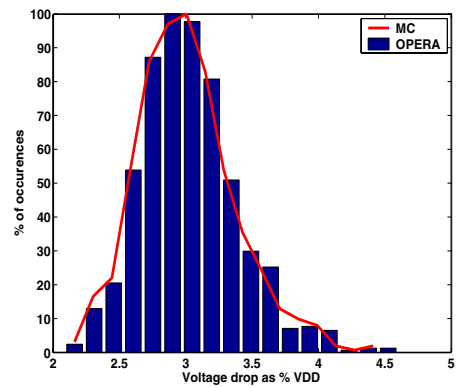

**Figure 1. Voltage distribution from OPERA and MC**

### 7. Conclusions

We presented a general framework to accurately compute the stochastic response of power grids in the presence of process variations. Using orthogonal polynomial expansions in a Hilbert space, we provide an explicit analytical representation of the stochastic response. The coefficients of the analytical expansion are determined precisely by utilising the orthogonality property of the polynomials of the expansion. Further, the expansion facilitates the direct and precise computation of the moments of the power grid response. We also show that our analysis becomes very simplified if we consider just the impact of leakage current variations on the power grid. We implemented the algorithm in a prototype software called OPERA and verified it extensively for many industrial grids. OPERA demonstrates very good accuracy when compared with the classical Monte Carlo simulations along with providing significant speed-ups up to two orders of magnitude.

| Size (# nodes) | Ave. % Error in $\mu$ | Max. % Error in $\mu$ | Ave. % Error in $\sigma$ | Max. % Error in $\sigma$ | $\pm 3\sigma$ variation (% of nominal drop $\mu_0$) | CPU time Monte (sec) | CPU time OPERA (sec) | Speedup |
|---|---|---|---|---|---|---|---|---|
| 19181 | 0.0155 | 0.0282 | 2.53 | 2.78 | $\pm 34$ | 1444.00 | 14.32 | 101 |
| 25813 | 0.0422 | 0.0838 | 3.41 | 3.84 | $\pm 33$ | 1565.30 | 77.93 | 20 |
| 34938 | 0.0204 | 0.5146 | 1.53 | 12.17 | $\pm 32$ | 1140.10 | 17.50 | 65 |
| 49262 | 0.1992 | 0.3713 | 6.73 | 7.37 | $\pm 37$ | 4777.87 | 178.52 | 27 |
| 62812 | 0.0680 | 0.1253 | 3.82 | 6.45 | $\pm 46$ | 1481.7 | 17.40 | 85 |
| 91729 | 0.0137 | 0.6037 | 3.28 | 18.03 | $\pm 30$ | 3172.67 | 25.50 | 124 |
| 351838 | 0.0926 | 0.1457 | 5.27 | 18.39 | $\pm 33$ | 109315 | 1050.72 | 104 |

**Table 1. Results for grids from OPERA and Monte Carlo simulations for order 2 expansion**

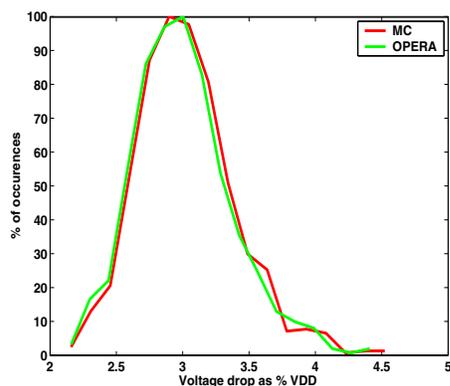

**Figure 2. Voltage distribution from OPERA and MC**

## 8. Acknowledgements

This work was carried out under the auspices of the National Science Foundation's Grant #CCR-0205227 and Grant #EEC-9523338. It was conducted at the National Science Foundation's State/Industry/University Cooperative Research Centers' (NSF-S/IUCRC) Center for Low Power Electronics (CLPE) which is supported by the NSF, the State of Arizona, and an industrial consortium. Any opinions, findings, and conclusions or recommendations expressed in this material are those of the authors and do not necessarily reflect the views of the National Science Foundation.